# The Private Classical Capacity of a Partially Degradable Quantum Channel

Laszlo Gyongyosi

[1]Quantum Technologies Laboratory, Department of Telecommunications
*Budapest University of Technology and Economics*
2 Magyar tudosok krt, Budapest, *H*-1117, Hungary
[2]Information Systems Research Group, Mathematics and Natural Sciences
*Hungarian Academy of Sciences*
Budapest, *H*-1518, Hungary

gyongyosi@hit.bme.hu

**Abstract**

For a partially degradable (PD) channel, the channel output state can be used to simulate the degraded environment state. The quantum capacity of a PD channel has been proven to be additive. Here, we show that the private classical capacity of arbitrary dimensional PD channels is equal to the quantum capacity of the channel and also single-letterizes. We prove that higher rates of private classical communication can be achieved over a PD channel in comparison to standard degradable channels.

## 1 Introduction

The existence of partially degradable (PD) channels has been recently investigated, and it was shown that PD channels exist in the world of those qudit channels that have entanglement-binding complementary channels [1]. The PD channel set involves conjugate-PD channels, which represent a subset in the conjugate degradable channels; this set was introduced by Bradler *et al.* [11]. The sets of *more capable* and *less noisy* quantum channels were defined by Watanabe [2], and it was further demonstrated that other, well-characterized channel sets could be defined beyond the degradable and conjugate-degradable channels. The quantum capacity [3], [6-8], [12-16] of some of these channels also single-letterizes, similar to the quantum capacity of PD channels [1]. However, at this point, the exact connection between these channel sets and PD channels is still open. It is clear, that a PD channel is equipped with the benefits of both the more capable and less noisy sets. However, in contrary to the more capable and less noisy sets, a PD channel can be also an anti-degradable one, since the structure of the complementary channels (the channel between the sender and the environment) differs at several points [1]. On the other hand, despite the lack of exact clarification of the connections between the sets, some properties of these channels can be exploited and combined together with the structural properties of PD channels to derive further valuable results regarding the capabilities of a PD channel.

It is also revealed that from the structure of a PD channel several important conclusions follow. First of all, these channels allow higher quantum data rates in comparison to standard, non-PD channels. Second, the PD channel structure allows similar benefits in the world of anti-degradable channels to the Hadamard channels in the degradable set, since the evaluation of its quantum capacity can easily be made in arbitrary dimensions. Similar to degradable channels [5], [9], the quantum capacity of a PD channel has been proven to be additive [1]. However, no similar results were obtained for the private classical capacity. Why is this so important for us? If we could prove that the private classical capacity of a PD channel is equal to its quantum capacity, then we could gain several benefits. First, we would know for sure that it single-letterizes, and second—and the reason that would probably have more significance for communication purposes—we would know that over a PD channel higher rates of private classical communication can be achieved in comparison to standard, non-PD degradable channels.

In this work, we prove these statements. First, we show that the private-classical capacity of a PD-channel single-letterizes; then, we conclude that it allows better rates of private classical communication in comparison to non-PD degradable channels.

This paper is organized as follows. In Section 2, we summarize the preliminary findings. In Sections 3 and 4, we present the main results. In Section 5, we conclude the paper.

## 2 Preliminaries

The *single-letter private classical capacity* (private information) can be expressed as

$$P^{(1)}(\mathcal{N}) = \max_{\text{all } p_i, \rho_{x'}} \left( I(A':B) - I(A':E) \right), \tag{1}$$

where $I(A':B)$ is the mutual information between the sender, Alice $(A')$ and the receiver, Bob $(B)$, while $I(A':E)$ stands for the mutual information between Alice and the eavesdropper, Eve $(E)$. The optimization is taken over $\{p_i, \rho_{x'}\}$, where $\rho_A^{x'} = \sum_{x'} p(x') \rho_{x'}$ is an arbitrary system over $\mathcal{H}_A$, with arbitrary probability distribution $p(x')$ on a finite set $A'$, $A' \in x'$, while $\rho_{x'}$ are not necessarily pure states [2]. The *private capacity* is defined as

$$P(\mathcal{N}) = \lim_{n \to \infty} \frac{1}{n} P^{(1)}(\mathcal{N}^{\otimes n}) = \lim_{n \to \infty} \frac{1}{n} \max_{\text{all } p_i, \rho_{x'}} \left( I(A':B) - I(A':E) \right). \tag{2}$$

The *single-letter quantum capacity* is defined as

$$Q^{(1)}(\mathcal{N}) = \max_{\text{all } p_i, \rho_x} I_{coh}(\rho, \mathcal{N}), \tag{3}$$

where $I_{coh}(\rho, \mathcal{N})$ is the coherent information, measured as $I_{coh}(\rho, \mathcal{N}) = \mathrm{S}(B) - \mathrm{S}(E)$, $B$ is the channel output, while $E$ is the environment's state, while $\mathrm{S}(\rho) = -Tr(\rho \log(\rho))$ is the von Neumann entropy. The maximization is made over input $\xi_A = \sum_x p(x) \rho_x$, $A \in x$, where $\rho_x = |\psi_x\rangle\langle\psi_x|$ are pure states, and $p(x)$ is a probability distribution over $A$ with $\dim|A| = \mathcal{H}_A$ [2]. The *quantum capacity* is defined as:

$$Q(\mathcal{N}) = \lim_{n \to \infty} \frac{1}{n} Q^{(1)}(\mathcal{N}^{\otimes n}) = \lim_{n \to \infty} \frac{1}{n} \max_{\text{all } p_i, \rho_x} I_{coh}(\xi_A, \mathcal{N}^{\otimes n}). \tag{4}$$

Further basic definitions will be revealed in the proof.

## 3 The private classical capacity single-letterizes

The main result on the private classical capacity of a PD channel is summarized in Theorem 1.

**Theorem 1.** *The private classical capacity of a PD channel single-letterizes and is equal to the quantum capacity of the channel.*

*Proof.*
The sketch of the proof as follows. The private classical and quantum capacities of a PD channel $\mathcal{N}_{PD}$ will be denoted by $P(\mathcal{N}_{PD})$, $Q(\mathcal{N}_{PD})$, respectively. First, we define an input system for which the general rule $P^{(1)}(\mathcal{N}_{PD}) \geq Q^{(1)}(\mathcal{N}_{PD})$ between the private information and the maximized coherent information holds, then we show that there exists another input system for which it turns to its opposite, i.e., $P^{(1)}(\mathcal{N}_{PD}) \leq Q^{(1)}(\mathcal{N}_{PD})$ holds. Then, by exploiting the properties of the complementary channel of a PD channel, we conclude that it finally has to led to $P(\mathcal{N}_{PD}) = Q(\mathcal{N}_{PD})$. The argument behind these statements is as follows.

Let us start from the beginning. For any channel $\mathcal{N}$, assuming input system $\rho_A$, the coherent information is expressed as

$$\begin{aligned} I_{coh}(\rho_A, \mathcal{N}) &= \mathcal{X}_{AB}(\mathcal{N}) - \mathcal{X}_{AE}(\mathcal{N}) \\ &= \mathcal{X}(\mathcal{N}_{AB}) - \mathcal{X}(\mathcal{N}_{AE}) \\ &= I(A:B) - I(A:E), \end{aligned} \tag{5}$$

where $\mathcal{X}_{AB}(\mathcal{N})$ is the Holevo information between Alice and Bob maximized over all $\{p_i, \rho_i\}$ input ensembles, $\mathcal{X}_{AE}(\mathcal{N})$ is the Holevo information between Alice and the environment, while $\mathcal{N}_{AB}$ and $\mathcal{N}_{AE}$ depict the logical channels between Alice ($A$) and Bob ($B$), and Alice and the environment $E$.

First, we have to take into consideration the properties of the complementary channel $\mathcal{N}_{AE}$ of a PD channel. It is already known that $\mathcal{N}_{AE}$ is an entanglement-binding channel that has no quantum capacity, but has positive private classical capacity, i.e., $Q(\mathcal{N}_{AE}) = 0$, $P(\mathcal{N}_{AE}) > 0$. It arises from the fact that an entanglement-binding channel generates a PPT (Positive Partial

Transpose) output system (i.e., a bound entangled system), which cannot be used for quantum communication, but allows private classical communication [1], [10].

Let $\mathcal{N}_{PD}$ be a PD channel with entanglement-binding complementary channel $\mathcal{N}_{AE}$, characterized by input system

$$\xi_A = \sum_x p(x)\rho_x, \tag{6}$$

where $\rho_x = |\psi_x\rangle\langle\psi_x|$ are pure states on $\mathcal{H}_A$, and $p(x)$ is a probability distribution over $A$ with $\dim|A| = \mathcal{H}_A$, $A \in x$, from which the relation follows for $\mathcal{N}_{AE}(\xi_A)$ of $\mathcal{N}_{PD}$:

$$\mathcal{N}_{AE}(\xi_A) : \{Q(\mathcal{N}_{AE}) = 0, P(\mathcal{N}_{AE}) > 0\}. \tag{7}$$

A PD channel $\mathcal{N}_{PD}$ with the *degraded* entanglement-binding complementary channel $\mathcal{N}_{AE'} = \mathcal{N}_{AE} \circ \mathcal{D}^{E \to E'}$, is characterized by the input system $\xi'_A$ as

$$\xi'_A = \sum_{x'} p(x')\rho_{x'}, \tag{8}$$

where $\rho_{x'}$ are not necessarily pure states on $\mathcal{H}_A$, and $p(x')$ is a probability distribution over the finite set $A'$, $A' \in x'$. From this, we have the following relation for $\mathcal{N}_{AE'}$ of $\mathcal{N}_{PD}$:

$$\mathcal{N}_{AE'}(\xi'_A) : \{Q(\mathcal{N}_{AE'}) = 0, P(\mathcal{N}_{AE'}) < P(\mathcal{N}_{AE})\}, \tag{9}$$

i.e., the amount of private classical information that is leaked to the Eve is lower for $\mathcal{N}_{AE'}$, in comparison to $\mathcal{N}_{AE}$. For inputs $\xi_A$ and $\xi'_A$, the relationships

$$\max_{\text{all } p_i, \rho_x} I_{coh}(\xi_A, \mathcal{N}_{PD}) \geq 0 : I(A:B) \geq I(A:E), \tag{10}$$

$$P^{(1)}(\mathcal{N}_{PD}) \geq 0 : I(A':B) \geq I(A':E) \tag{11}$$

follow for the maximized coherent information and the private information, from which we immediately conclude that for $\xi_A$ and $\xi'_A$,

$$P^{(1)}(\mathcal{N}_{PD}) \geq Q^{(1)}(\mathcal{N}_{PD}). \tag{12}$$

Since for a PD channel the quantum capacity single-letterizes [1], it leads to

$$P(\mathcal{N}_{PD}) = P^{(1)}(\mathcal{N}_{PD}) \geq Q(\mathcal{N}_{PD}) = Q^{(1)}(\mathcal{N}_{PD}). \tag{13}$$

To step further in the proof, we define an arbitrary input system $\rho_A$ on $\mathcal{H}_A$ as

$$\rho_A \equiv \sum_{x,x'} p(x')p(x',x|x')|\psi_{x,x'}\rangle\langle\psi_{x,x'}|, \tag{14}$$

where $x$ and $x'$ are random variables over set $A \times A'$, $p(x')$ and $p(x')p(x',x|x')$ are the probability distributions on set $A \times A'$, while $|\psi_{x,x'}\rangle$ are pure states on $\mathcal{H}_A$ [2]. The probability distribution $p(x')$ is arbitrary. The system defined in (14) can be rewritten as

$$\rho_A = \sum_{x'} p(x')\rho_A^{x'}, \tag{15}$$

where $\rho_A^{x'} = \sum_x p(x',x|x')|\psi_{x,x'}\rangle\langle\psi_{x,x'}|$ is arbitrary [2], and the density matrix $\rho_A^{x'}$ refers to a system on $\mathcal{H}_A$. Assuming $\rho_A$ defined in (15), the maximized coherent information of $\mathcal{N}_{PD}$ is as follows

$$Q^{(1)}\left(\mathcal{N}_{PD}\right) = \max_{\text{all } p_i, \rho_A^{x'}} I_{coh}\left(\rho_A, \mathcal{N}_{PD}\right) = \max_{\text{all } p_i, \rho_A^{x'}} \left(I\left(A:B\right) - I\left(A:E\right)\right), \tag{16}$$

where $E$ is the environment state; while for the private information, (15) leads to

$$\begin{aligned} P^{(1)}\left(\mathcal{N}_{PD}\right) &= \max_{\text{all } p_i, \rho_A^{x'}} \left(I\left(A':B\right) - I\left(A':E\right)\right) \\ &= \max_{\text{all } p_i, \rho_A^{x'}} \left(I\left(A:B\right) - I\left(A:E\right) - I\left(A:B|A'\right) + I\left(A:E|A'\right)\right), \end{aligned} \tag{17}$$

from which the relationship between (16) and (17) is

$$\max_{\text{all } p_i, \rho_A^{x'}} I_{coh}\left(\rho_A, \mathcal{N}_{PD}\right) \geq \max_{\text{all } p_i, \rho_A^{x'}} \left(I\left(A':B\right) - I\left(A':E\right)\right). \tag{18}$$

For (15), the following connection holds between the single-letter capacities:

$$P^{(1)}\left(\mathcal{N}_{PD}\right) = Q^{(1)}\left(\mathcal{N}_{PD}\right) - \eta, \tag{19}$$

where $\eta \geq 0$ is a weighting parameter, defined as

$$\eta = \max_{\text{all } p_i, \rho_A^{x'}} \left(I\left(A:B|A'\right) - I\left(A:E|A'\right)\right), \tag{20}$$

where $A' = x'$. In other words, parameter $\eta$ is the difference of private information and coherent information functions obtained in (16) and (17). For $Q^{(1)}\left(\mathcal{N}_{PD}\right) \geq 0$

$$Q^{(1)}\left(\mathcal{N}_{PD}\right) \geq 0 : I\left(A:B\right) \geq I\left(A:E\right), \tag{21}$$

holds, while for the non-negative private information, we have

$$P^{(1)}\left(\mathcal{N}_{PD}\right) \geq 0 : I\left(A:B\right) - I\left(A:B|A'\right) \geq I\left(A:E\right) - I\left(A:E|A'\right). \tag{22}$$

The input system $\rho_A$ (defined in (14)) maximizes the coherent information since $\left\{\left|\psi_{x,x'}\right\rangle\right\}$ are pure states, hence the following relationship holds between $P^{(1)}$ and $Q^{(1)}$:

$$\begin{aligned} P^{(1)}\left(\mathcal{N}_{PD}\right) &= \max_{\text{all } p_i, \rho_A^{x'}} \left(I\left(A':B\right) - I\left(A':E\right)\right) \leq \\ Q^{(1)}\left(\mathcal{N}_{PD}\right) &= \max_{\text{all } p_i, \rho_A^{x'}} I_{coh}\left(\rho_A, \mathcal{N}_{PD}\right) = \max_{\text{all } p_i, \rho_A^{x'}} \left(I\left(A:B\right) - I\left(A:E\right)\right), \end{aligned} \tag{23}$$

and by exploiting the fact that the quantum capacity of a PD-channel single-letterizes [1] it leads to

$$P\left(\mathcal{N}_{PD}\right) = P^{(1)}\left(\mathcal{N}_{PD}\right) \leq Q\left(\mathcal{N}_{PD}\right) = Q^{(1)}\left(\mathcal{N}_{PD}\right). \tag{24}$$

From the inequalities (13) and (24), the following conclusion can be made for the private information and the coherent information of a PD channel for all $\rho_i$:

$$P^{(1)}\left(\mathcal{N}_{PD}\right) = \max_{\text{all } p_i, \rho_i} I_{coh}\left(\rho_i, \mathcal{N}_{PD}\right). \tag{25}$$

Since the quantum capacity of a PD channel single-letterizes [1], it leads to the following conclusion regarding the private classical capacity of a PD channel:

$$P^{(1)}\left(\mathcal{N}_{PD}\right) = P\left(\mathcal{N}_{PD}\right) = Q^{(1)}\left(\mathcal{N}_{PD}\right) = Q\left(\mathcal{N}_{PD}\right), \tag{26}$$

which completes the proof.

These results conclude the proof of Theorem 1.

∎

# 4 Rate of private classical communication

The main result on the rate of private classical communication over PD channel is summarized in Theorem 2.

**Theorem 2.** *The maximal rate of private classical communication over a PD channel is higher than that of standard degradable channels.*

*Proof.*

The results strictly follow from the results obtained in the previous section. Since then, we have found that the private classical capacity is, indeed, equal to the quantum capacity of the channel. For a PD channel, it was already shown that it allows better quantum data rates in comparison to non-PD channels, thanks to the partial simulation property that benefit arises from the structure of a PD channel [1]. On the other hand, for a degradable channel, it was shown that its private classical capacity equals to the quantum capacity [5], [9]. Thanks to the structural properties of the PD set, higher quantum data rates can be obtained in comparison to standard, non-PD degradable channels [1]. By combining the results of [1] on the quantum capacity of a PD channel, with the results obtained in Theorem 1, it will follow that a PD channel allows higher rates of private classical communication.

For a degradable channel, it is known that $P^{(1)}(\mathcal{N}_D) = P(\mathcal{N}_D) = Q^{(1)}(\mathcal{N}_D) = Q(\mathcal{N}_D)$. Here, we show that for the $R_P$ rate of private classical communication for a degradable $\mathcal{N}_D$ and a PD channel $\mathcal{N}_{PD}$ the relation $R_P(\mathcal{N}_{PD}) \geq R_P(\mathcal{N}_D)$ holds. We will find that for a PD channel, the amount of classical (non-private) information that is leaked to Eve is lower in comparison to non-PD degradable channels, and —thanks to Theorem 1, we already know that it is, in fact, equal to the amount of classical information that is leaked to the environment. What does it follow from this result? An important corollary. Namely, it leads to $R_P(\mathcal{N}_{PD}) = R_Q(\mathcal{N}_{PD})$, where $R_Q(\mathcal{N}_{PD})$ is the rate of quantum communication over a PD channel.

To see it, let us allow to rewrite the previously introduced capacity formulas. First, we exploit the fact that $P(\mathcal{N}_D)$ and $P(\mathcal{N}_{PD})$ of a $\mathcal{N}_D$ and $\mathcal{N}_{PD}$ channel can be expressed as follows:

$$\begin{aligned} P(\mathcal{N}_D) &= \max_{\text{all } p_i,\rho_i} \left( \mathcal{X}_{AB}(\mathcal{N}_D) - \mathcal{X}_{AE}(\mathcal{N}_D) \right) \\ &= \max_{\text{all } p_i,\rho_i} \left( \mathcal{X}(\mathcal{N}^D_{AB}) - \mathcal{X}(\mathcal{N}^D_{AE}) \right), \end{aligned} \quad (27)$$

and

$$\begin{aligned} P(\mathcal{N}_{PD}) &= \max_{\text{all } p_i,\rho_i} \left( \mathcal{X}_{AB}(\mathcal{N}_{PD}) - \mathcal{X}_{AE'}(\mathcal{N}_{PD}) \right) \\ &= \max_{\text{all } p_i,\rho_i} \left( \mathcal{X}(\mathcal{N}^{PD}_{AB}) - \mathcal{X}(\mathcal{N}^{PD}_{AE'}) \right), \end{aligned} \quad (28)$$

where $\mathcal{N}_{AB}$ and $\mathcal{N}_{AE}$ are the logical channels between Alice and Bob, and Alice and Eve, respectively. For a PD channel, the complementary channel is $\mathcal{N}_{AE'} = \mathcal{N}_{AE} \circ \mathcal{D}^{E \to E'}$, where $\mathcal{N}_{AE}$ is the entanglement-binding channel, while $\mathcal{D}^{E \to E'}$ is a degradation map on $E$. Let us to clarify this result. The quantities $\mathcal{X}(\mathcal{N}_{AE}^{D})$ and $\mathcal{X}(\mathcal{N}_{AE} \circ \mathcal{D}^{E \to E'})$ are, in fact, measure the classical information that is leaked to Eve (i.e., to the environment, see (26)) during the transmission via the complementary channels $\mathcal{N}_{AE}^{D}$ and $\mathcal{N}_{AE'}^{PD}$. The Holevo quantities of the $\mathcal{N}_{AB}^{D}$ and $\mathcal{N}_{AE}^{D}$ logical channels for a degradable channel are

$$\mathcal{X}(\mathcal{N}_{AB}^{D}) = S(\mathcal{N}_{AB}^{D}(\rho_{AB})) - \sum_i p_i S(\mathcal{N}_{AB}^{D}(\rho_i)), \tag{29}$$

$$\mathcal{X}(\mathcal{N}_{AE}^{D}) = S(\mathcal{N}_{AE}^{D}(\rho_{AE})) - \sum_i p_i S(\mathcal{N}_{AE}^{D}(\rho_i)), \tag{30}$$

where $\rho_{AB} = \sum_i p_i \rho_i$ and $\rho_{AE} = \sum_i p_i \rho_i$ are the average states.

For a $\mathcal{N}_{PD}$ channel, with the logical channels $\mathcal{N}_{AB}^{PD}$ and $\mathcal{N}_{AE'}^{PD}$ we write

$$\mathcal{X}(\mathcal{N}_{AB}^{PD}) = S(\mathcal{N}_{AB}^{PD}(\rho_{AB})) - \sum_i p_i S(\mathcal{N}_{AB}^{PD}(\rho_i)), \tag{31}$$

$$\mathcal{X}(\mathcal{N}_{AE'}^{PD}) = S(\mathcal{N}_{AE'}^{PD}(\rho_{AE})) - \sum_i p_i S(\mathcal{N}_{AE'}^{PD}(\rho_i)). \tag{32}$$

The rate improvement is a corollary that follows from the structural differences of the complementary channels $\mathcal{N}_{AE}^{D}$ and $\mathcal{N}_{AE'}^{PD}$.

The difference of the Holevo information of the complementary channels of a $\mathcal{N}_D$ and $\mathcal{N}_{PD}$ channel can be quantified as

$$\Delta = \max_{\text{all } p_i, \rho_i} \left( \mathcal{X}(\mathcal{N}_{AE}^{D}) - \mathcal{X}(\mathcal{N}_{AE'}^{PD}) \right). \tag{33}$$

To see it, we express the quantities $\mathcal{X}(\mathcal{N}_{AE}^{D})$ and $\mathcal{X}(\mathcal{N}_{AE'}^{PD})$ as

$$\mathcal{X}(\mathcal{N}_{AE}^{D}) = I(A' : E), \tag{34}$$

$$\mathcal{X}(\mathcal{N}_{AE'}^{PD}) = I(A : E) - \Delta, \tag{35}$$

where variable $A'$ characterizes the $\mathcal{N}_D$ channel with complementary channel $\mathcal{N}_{AE}^{D}$, while variable $A$ identifies the channel $\mathcal{N}_{PD}$ with $\mathcal{N}_{AE'}^{PD}$, hence

$$P(\mathcal{N}_{PD}) - P(\mathcal{N}_D) = \max_{\text{all } p_i, \rho_i} \left( \mathcal{X}(\mathcal{N}_{AE}^{D}) - \mathcal{X}(\mathcal{N}_{AE'}^{PD}) \right) = \Delta, \tag{36}$$

where $\Delta \geq 0$. The quantities $\mathcal{X}(\mathcal{N}_{AB}^{D})$ and $\mathcal{X}(\mathcal{N}_{AB}^{PD})$ are

$$\mathcal{X}(\mathcal{N}_{AB}^{D}) = I(A' : B), \tag{37}$$

$$\mathcal{X}(\mathcal{N}_{AB}^{PD}) = I(A : B), \tag{38}$$

These clearly demonstrate that the improvement in the private classical capacity is

$$\Delta = \max_{\text{all } p_i, \rho_i} \left( I(A : B | A') - I(A : E | A') \right). \tag{39}$$

This leads to the following private classical capacity for a PD channel

$$\begin{aligned}
P(\mathcal{N}_{PD}) &= P(\mathcal{N}_D) + \Delta \\
&= \max_{\text{all } p_i, \rho_i} \left( \mathcal{X}(\mathcal{N}_{AB}^D) - \left( \mathcal{X}(\mathcal{N}_{AE}^D) - \mathcal{X}(\mathcal{N}_{AE}^D) + \mathcal{X}(\mathcal{N}_{AE'}^{PD}) \right) \right) \\
&= \max_{\text{all } p_i, \rho_i} \left( \mathcal{X}(\mathcal{N}_{AB}^{PD}) - \mathcal{X}(\mathcal{N}_{AE'}^{PD}) \right) \\
&= \max_{\text{all } p_i, \rho_i} \left( I(A:B) + I(A:B|A') - I(A:E) - I(A:E|A') \right) \\
&= \max_{\text{all } p_i, \rho_i} \left( I(A:B) - I(A:E) + \left( I(A:B|A') - I(A:E|A') \right) \right) \\
&= \max_{\text{all } p_i, \rho_i} \left( I(A:B) - \left( I(A:E) - \Delta \right) \right),
\end{aligned} \quad (40)$$

from which the conclusion

$$P(\mathcal{N}_{PD}) \geq P(\mathcal{N}_D) \quad (41)$$

follows for any $\Delta \geq 0$. From the results obtained in [1], we know that for the quantum capacity of a $\mathcal{N}_D$ and $\mathcal{N}_{PD}$ channel, the relationship

$$Q(\mathcal{N}_{PD}) = Q(\mathcal{N}_D) + \Delta, \quad (42)$$

holds, while from the proof of Theorem 1, we know that the quantum capacity is, in fact, equal to the private capacity, from which statements, we finally get

$$\begin{aligned}
P(\mathcal{N}_{PD}) &= P^{(1)}(\mathcal{N}_{PD}) \\
&= P^{(1)}(\mathcal{N}_D) + \Delta \\
&= \max_{\text{all } p_i, \rho_i} I_{coh}(\rho_A, \mathcal{N}_{PD}) \\
&= \max_{\text{all } p_i, \rho_i} I_{coh}(\rho_A, \mathcal{N}_D) + \Delta \\
&= Q(\mathcal{N}_{PD}) \\
&= Q(\mathcal{N}_D) + \Delta \\
&= P(\mathcal{N}_D) + \Delta.
\end{aligned} \quad (43)$$

From the results obtained in (43), it follows that by using capacity-achieving codes, for the rates of private classical communication the following connection can be achieved:

$$R_P(\mathcal{N}_{PD}) = R_Q(\mathcal{N}_{PD}) = R_P(\mathcal{N}_D) + \Omega, \quad (44)$$

where $0 \leq \Omega \leq \Delta$ is the *rate improvement*, which can be achieved by the given channel coding scheme (limited to $\Delta$). For the rate of private classical communication over a $\mathcal{N}_{PD}$ and $\mathcal{N}_D$ channel, the relationship

$$R_P(\mathcal{N}_{PD}) \geq R_P(\mathcal{N}_D) \quad (45)$$

clearly follows.

At this point we summarize the results. We have confirmed that for a PD channel, the same improvement can be realized in the transmission rate of private classical information than for the quantum data rate of the channel. This result leads to higher rates of private classical communication in comparison to standard, non-PD degradable channels.

These results conclude the proof of Theorem 2.

∎

# 5  Conclusions

We showed the partial degradability property can be further exploited and the benefits can be extended for the transmission of private classical information. We have proven that the private classical capacity of a PD channel single-letterizes and is equal to the quantum capacity of that channel. We also revealed that higher rates of private classical communication can be achieved over a PD channel in comparison to standard, non-PD degradable channels. These results demonstrate that the structural benefits of a PD-channel can be further exploited and allowed to exceed the current boundaries on the rates of achievable private classical communication over standard degradable channels.

# Acknowledgements

The author would like to thank Shun Watanabe for suggesting Ref [2]. The results discussed above are supported by the grant COST Action MP1006.# References

[1]  L. Gyongyosi. The structure and quantum capacity of a partially degradable quantum channel. *IEEE Access, arXiv:1304.5666,* (2014).

[2]  S. Watanabe. Private and quantum capacities of more capable and less noisy quantum channels. *Physical Review A 85, 012326* (2012).

[3]  S. Imre and L. Gyongyosi. *Advanced Quantum Communications - An Engineering Approach.* Wiley-IEEE Press (New Jersey, USA), (2012).

[4]  L. Gyongyosi. Quantum information transmission over a partially degradable channel, *IEEE Access, arXiv:1303.0606* (2014).

[5]  T. S. Cubitt, M. B. Ruskai, and G. Smith. The structure of degradable quantum channels. *Journal of Mathematical Physics, 49:102104,* (2008).

[6]  S. Lloyd. Capacity of the noisy quantum channel. *Physical Review A, 55(3):1613,* (1997).

[7]  P. W. Shor. The quantum channel capacity and coherent information. *In lecture notes, MSRI Workshop on Quantum Computation,* (2002).

[8]  I. Devetak. The private classical capacity and quantum capacity of a quantum channel. *IEEE Trans. Inf. Theory, vol. 51, pp. 44–55.* quant-ph/0304127, (2005).

[9]  I. Devetak and P. W. Shor. The capacity of a quantum channel for simultaneous transmission of classical and quantum information. *Commun. Math. Phys. 256, 287,* (2005).

[10]  P. Horodecki, M. Horodecki, and R. Horodecki. Binding entanglement channels. *J. Mod. Opt., vol. 47,* pp. 347–354, 2000.

[11]  K. Bradler, N. Dutil, P. Hayden, A. Muhammad. Conjugate degradability and the quantum capacity of cloning channels. *Journal of Mathematical Physics 51,* 072201 (2010).

[12]  L. Hanzo, H. Haas, S. Imre, D. O'Brien, M. Rupp, L. Gyongyosi. Wireless Myths, Re-alities, and Futures: From 3G/4G to Optical and Quantum Wireless, *Proceedings of the IEEE,* Volume: 100, Issue: Special Centennial Issue, pp. 1853-1888. (2012).9